\documentclass[summary]{URSIGASS2020}
\usepackage[bookmarks=false]{hyperref}
\usepackage{graphicx,color,epsfig}
\usepackage{amssymb,epsf,amsmath}

\title{Scaling performance of the SAGECal calibration package: from LOFAR to SKA}

\author{H. Spreeuw*\affref{ref1},  S. Yatawatta\affref{ref2},
  B. Van  Werkhoven\affref{ref1}, and F. Diblen\affref{ref1}}

\affiliation{%
  \aff{ref1}{Netherlands eScience Center, Science Park 140, Amsterdam, The Netherlands, https://www.esciencecenter.nl}
  \aff{ref2}{ASTRON, The Netherlands Institute for Radio Astronomy, Dwingeloo, The Netherlands, https://www.astron.nl/}
}

\def\mnras{Monthly Notices of the Royal Astronomical Society} % not picked up in bibtex

\begin{document}

\maketitle

\begin{abstract}
  This decade, the Square Kilometre Array (SKA) will perform its first observations. Preparations for building dishes, antennas, correlators and infrastructure are well underway. Concurrently, software for the processing of SKA observations is developed at a number of levels. At a more basic level there are the telescope monitoring and control systems and also the correlator software. On top of that, in order to deliver science ready data products, software pipelines are needed for radio frequency interference (RFI) mitigation, averaging, calibration and imaging. Here, we focus on the SAGECal calibration package, in particular on the times needed to obtain calibration solutions. This is an important aspect, since this package is now used for the Epoch of Reionization (EoR) Key Science Project of LOFAR, but will also have to run optimally on SKA1 LOW. In terms of number of stations used for observing this amounts to a factor 10 increase, from 51 to 512 stations. Consequentially, the disk space needed to store an observation will increase by a factor 100, provided the number of frequency channels remains the same. In this paper we investigate the scaling behaviour of SAGECal, whose runtimes should ideally scale linearly with the number of stations. We also explain the algorithms inside SAGECal and use them to explain its scaling behaviour.
\end{abstract}

\section{Introduction}

A number of radio telescopes with many low-gain elements have recently been deployed, designed and/or funded. However, the history of such telescopes goes back to the Mills Cross Telescopes in Australia \footnote{\url{https://www.atnf.csiro.au/news/newsletter/jun02/Flowering_of_Fleurs.htm}} and the U.S. \footnote{\url{https://radiojove.gsfc.nasa.gov/library/sci_briefs/discovery.html}} in the 1950s and the Clark Lake TPT telescope in the 1960s ~\cite{1982ApJS...50..403E}. Prior to that, in the 1940s, astronomical radio interferometry had begun in Australia with a single radar antenna, a sea-cliff interferometer ~\cite{1991ASPC...19..132S}.
Unfortunately for the Clark Lake Telescope, ionospheric effects limited its calibration to short baselines ~\cite{2005ASPC..345..114K}. Self-calibration, a new calibration technique to mitigate these effects, came too late and the Clark Lake TPT could in no way compete with the scientific potential of the VLA, which operated at higher frequencies. These authors ~\cite{2005ASPC..345..114K} explain why two telescopes with a factor of 10 difference in angular resolution will have a factor of 100 difference in sensitivity. This sophisticated instrument, with a collecting area about 3 times larger than the VLA, closed down in the early 1990s, which marked a period of several decades with little progress in low frequency radio astronomy. This only came to an end with the operation of the Long Wavelength Array (LWA\footnote{\url{http://www.phys.unm.edu/~lwa/index.html}}) and the LOw Frequency Array (LOFAR\footnote{\url{https://www.astron.nl/telescopes/lofar}}).
With the advent of these two large telescopes, quite different problems, related to required disk space and compute power arose. These originate from unprecedented available bandwidth and frequency resolution, but also from the large numbers of stations acquiring data for each observation. Since the size, in disk space, of each observation scales with the number of baselines, it scales quadratically with the number of stations. 
LWA and LOFAR are regarded as pathfinders for the Square Kilometre Array (SKA)\footnote{\url{https://www.skatelescope.org/precursors-pathfinders-design-studies/}}. The SKA will operate at two sites, in Australia at low (SKA1 LOW) and in South-Africa at intermediate (SKA1 MID) radio frequencies. For SKA1 LOW 512 stations will be built \footnote{\url{https://www.skatelescope.org/technical/info-sheets/}}, which amounts to data sizes a factor 100 larger than LOFAR - 51 Low Band stations as of mid 2019 \footnote{\url{http://old.astron.nl/radio-observatory/astronomers/technical-information/lofar-technical-information}} - if the number of frequency channels would be comparable. 
For the reduction pipelines that turn the raw observational data into science ready data products such large amounts of data pose severe challenges. Apart from the (temporary) disk space needed, it is essential that the throughput of observational data keeps pace with its collection. The reduction pipelines will perform RFI mitigation, averaging, calibration and imaging. Of these four tasks, calibration is the most compute intensive, and we focus on that aspect in this paper, by investigating the performance of the SAGECal calibration package for numbers of stations comparable to LOFAR up to the number of stations expected for SKA1 LOW. Also, we investigate its dependence on the number of sources in the sky model. In reality, accurate calibration will involve a number of self-calibration loops, which include not only calibration, but also imaging, but we will neglect the latter for now.

\section{Description of the SAGECal calibration package}
SAGECal is an amalgamation of several key algorithms for optimal performance and accuracy in calibration. We briefly discuss these algorithms in the following and highlight their unique characteristics.
\begin{itemize}
\item Expectation maximization (EM) and space alternating generalized expectation maximization (SAGE) algorithms \cite{Fess94}: These algorithms enable us to decompose calibration along multiple directions in the sky into a set of uni-directional calibration problems \cite{Kaz2}. Therefore, we get linear scaling with the number of directions being calibrated.
\item Expectation-conditional maximization either (ECME) algorithm \cite{Liu1995}: We use this algorithm to minimize a Student's T cost function \cite{Kaz3} in robust calibration.
\item Limited memory, Broyden Fletcher Goldfarb Shanno (LBFGS) algorithm \cite{Fletcher}: By approximating the (inverse) Hessian using a set of vectors, we can calibrate with a large number of unknowns, without running out of memory. Moreover, by using a stochastic LBFGS algorithm \cite{DSW2019}, we can calibrate data at the highest resolution.
\item Riemannian trust region (RTR) algorithm \cite{RTR}: Combined with the truncated conjugate gradient method \cite{Trunc1,Trunc2}, we get a solver that scales linearly in memory usage with the number of stations \cite{SS5} unlike quadratic scaling of other Newton methods. Moreover, because we calibrate on a Riemannian quotient manifold, we get faster convergence compared to Euclidean space calibration.
\item Alternating direction method of multipliers (ADMM) \cite{boyd2011}: We can calibrate data taken over a wide bandwidth and also data that are stored across a compute cluster by formulating it as a distributed consensus optimization problem \cite{DCAL,DMUX}.
\end{itemize}

All the above algorithms have been accelerated using GPUs \cite{InPar,Spreeuw2019}. We show the scaling of the workhorses of SAGECal, i.e., the optimization algorithms RTR and LBFGS, using simulated data. We simulate mock radio telescopes having the number of stations $N$ from 64 (LOFAR) to 1024 (super SKA). The calibration is performed along one direction only because the linear scaling with more than one direction is well illustrated in our previous work \cite{InPar,Spreeuw2019}.

\begin{figure}[htb]
\begin{minipage}[b]{0.98\linewidth}
\centering
 \centerline{\epsfig{figure=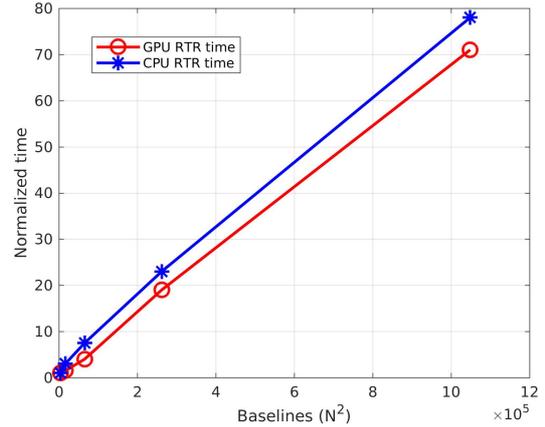,width=8.0cm}}
\end{minipage}
\caption{The scaling of RTR algorithm with $N^2$. Compared to LOFAR, SKA will require about $\times 20$ more time.\label{time_rtr}}
\end{figure}

\begin{figure}[htb]
\begin{minipage}[b]{0.98\linewidth}
\centering
 \centerline{\epsfig{figure=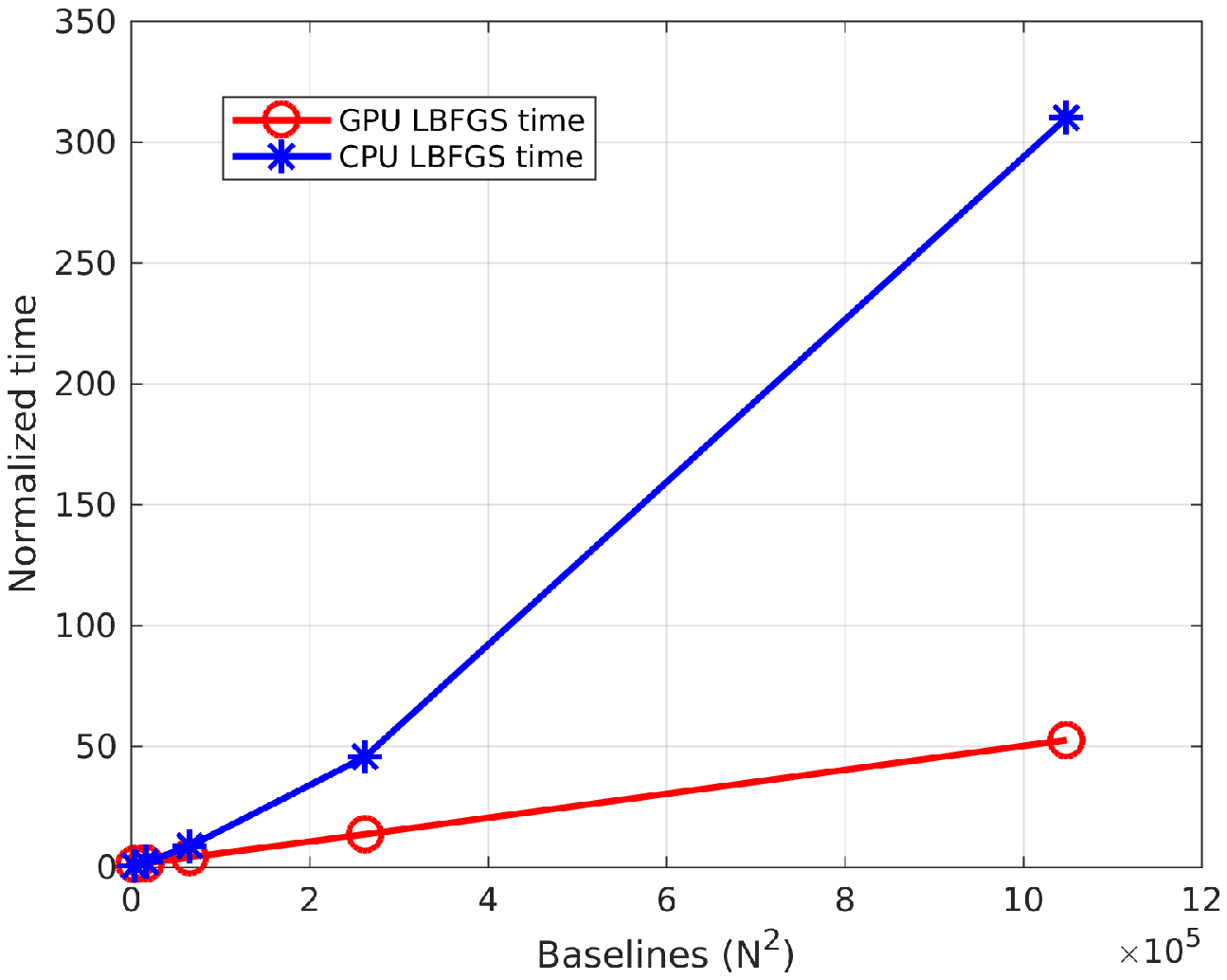,width=8.0cm}}
\end{minipage}
\caption{The scaling of LBFS algorithm with $N^2$. Compared to LOFAR, SKA will require about $\times 20$ ($\times 50$ for CPU) more time.\label{time_lbfgs}}
\end{figure}

\begin{figure}[htb]
\begin{minipage}[b]{0.98\linewidth}
\centering
 \centerline{\epsfig{figure=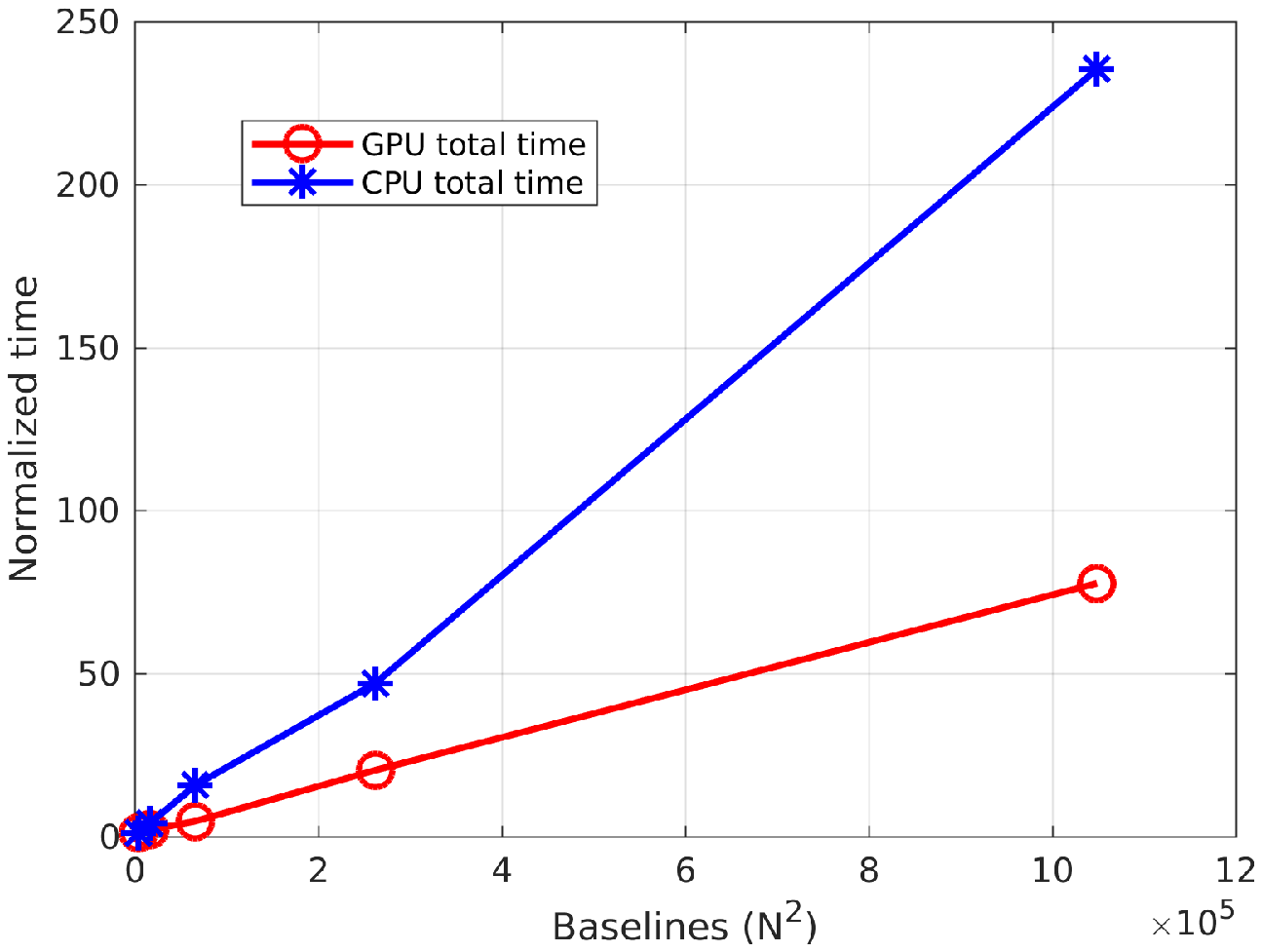,width=8.0cm}}
\end{minipage}
\caption{The scaling of both RTR and LBFGS algorithms with $N^2$. Compared to LOFAR, SKA will require about $\times 20$ ($\times 50$ for CPU) more time.\label{time_total}}
\end{figure}

For the comparison, we use a 12 core Intel Xeon E5-2670 CPU and an NVIDIA Tesla K40c GPU. We plot the time taken by RTR and LBFGS as well as the total time taken to calibrate $10$ time samples. We plot the time taken (normalized by the GPU time for $N=64$) against $N^2$ (instead of $N$) in Figs. \ref{time_rtr},\ref{time_lbfgs} and \ref{time_total}. This is to illustrate the scaling with the number of baselines which is proportional to $N^2$.

From Fig. \ref{time_rtr}, we see that the performance of RTR algorithm is similar in both GPU and CPU versions. The CPU is not overwhelmed while running the RTR algorithm and can keep up with the GPU. In contrast, we see a significant overhead in the CPU version of LBFGS in Fig. \ref{time_lbfgs}. The total time in Fig. \ref{time_total} is dominated by the LBFGS algorithm. Overall, we clearly see the scaling that is dependent on the data, which is in turn dependent on the number of baselines. Using mini-batches of data as in \cite{DSW2019} will improve this and we pursue this as future work. Note also that we have only shown the scaling of the optimization algorithms and for full calibration, additional time is required for model computation and data reading and writing \cite{Spreeuw2019}.
\section{Conclusions}
We have shown the scaling of the core optimization algorithms used in SAGECal. With GPU acceleration, we require about $\times 20$ more computations when we scale from LOFAR to SKA. Considering that there is $\times 8$ increase in the number of stations and $\times 64$ increase in the number of baselines, we see that the scaling is superlinear.
\section{Acknowledgements}

This work is supported by Netherlands eScience Center (project DIRAC, grant 27016G05).

%\bibliographystyle{IEEE}
%\bibliography{references}

\end{document}